# TOWARDS STANDARDIZED REGULATIONS FOR BLOCK CHAIN SMART CONTRACTS: INSIGHTS FROM DELPHI AND SWARA ANALYSIS

Shahin Heidari[1], Shannon Hashemi[2], Mohammad-Soroush Khorsand[3],
Alireza Daneshfar[4] and Seyedalireza Jazayerifar[5]

**ABSTRACT**

The rise of digital currency and the public ledger Block Chain has led to the development of a new type of electronic contract known as "smart contracts." For these contracts to be considered valid, they must adhere to traditional contract rules and be concluded without any impediments. Once written, encrypted, and signed, smart contracts are recorded in the Block Chain Ledger, providing transparent and secure record-keeping. Smart contracts offer several benefits, including their ability to execute automatically without requiring human intervention, their provision of public visibility of contract provisions on the Block Chain, their avoidance of financial crimes like Money Laundering, and their prevention of contract abuses. However, disputes arising from smart contracts still require human intervention, presenting unique challenges in enforcing these contracts, such as evidentiary issues, enforceability of waivers of defenses, and jurisdictional and choice-of-law considerations. Due to the novel nature of smart contracts, there are currently no standardized regulations that apply to them. Countries that have approved them have turned to customary law to legitimize their use. The Delphi method was used to identify critical success factors for applying blockchain transactions in a manufacturing company. Stepwise Weight Assessment Ratio Analysis (SWARA) was then utilized to determine the most influential factors. The proposed methodology was implemented, and results show that the most influential factors for the successful application of blockchain transactions as smart contracts in a manufacturing company are: turnover, the counter argument, vision, components for building, and system outcome quality. Conversely, connections with government entities and subcontractors, and the guarantee of quality have the least influence on successful implementation. These findings can contribute to the development of a legal framework for smart contracts in a manufacturing company.

**Keywords**: smart contracts; bitcoin; block chain; technology; legal rules.

## 1. INTRODUCTION

The banking and financial press have become increasingly familiar with the terms bitcoin, blockchain, and smart contracts (Cong & He, 2019; Idelberger et al., 2016). Bitcoin's underlying blockchain technology is utilized to address a number of perplexing issues (Watanabe et al., 2016; Hewa et al., 2021), including the need to lower transaction fees, increase processing speeds, broaden access to financial services, and give customers more agency. The original goals of blockchain development were to lower transaction costs, increase speed and efficiency in commercial transactions, and do away with the necessity for a third party to execute legal contracts (Wang et al., 2019; Ante, 2021; Laube and Sasani, 2020; Abbasi et al., 2022). To ensure the creation and implementation of the single platform, the World Economic Forum believes that smart contracts using blockchain technology can be put into place. Because to the decreased need for human intervention, financial agreements can be put into action more quickly, which in turn speeds up business operations (Khan et al., 2021; Sillaber & Waltl, 2017; Ibrahim and Sasani, 2021). Others have opined that if smart contracts can be executed on the blockchain, they won't need to be enforced through the courts (Sillaber & Waltl, 2017; Gatteschi et al., 2018). By facilitating the use of smart contracts, many previously unachievable objectives and forecasts have become a reality. Some people don't take this problem seriously and claim that smart contracts, smart contracts, and contracts don't exist. If smart contracts gain traction as useful corporate tools, more legal and regulatory guidelines may be necessary to limit unintended consequences and unleash their full potential. They have proven their legitimacy (Alipour and Charandabi, 2023, Singh et al., 2020; Kosba et al., 2016).

Despite the potential benefits offered by smart contracts, there are still several research gaps that need to be addressed. Firstly, there is a lack of standardized regulations that apply to smart contracts due to their novel nature, making it difficult for businesses and individuals to navigate the legal landscape. This has resulted in countries relying on

1. MBA Pompea College of Business, University of New Haven, USA.
2. Ernest C. Trefz School of Business, University of Bridgeport, USA.
3. Department of Accounting, Science and Research Branch, Islamic Azad University, Iran.
4. Ph.D., University of New Haven, West Haven, USA.
5. Department of Marketing, Walton College of Business, University of Arkansas, Fayetteville, Arkansas, USA

customary law to legitimize their use, creating inconsistencies in their application across different jurisdictions. Secondly, while smart contracts offer the potential for automated execution and increased transparency, disputes arising from these contracts still require human intervention to resolve, presenting unique challenges in enforcement. Finally, there is a need for further research into the technical challenges associated with implementing smart contracts in different industries, including manufacturing, to fully unlock their transformative potential. Addressing these research gaps will be crucial to realizing the full benefits of smart contracts and ensuring their successful integration into various industries, including manufacturing.

The sections of this article are as follows: The second section of this report provides a comprehensive background on blockchain technology and smart contracts, explaining their underlying concepts and how they work. In the final section, we delve deeper into the specifics of implementing this cutting-edge innovation, discussing potential technical challenges that may arise and proposing solutions to overcome them. Moreover, the report examines legal and regulatory hurdles that must be addressed before smart contracts can achieve widespread recognition and adoption. We emphasize the need for legislative amendments to accelerate the realization of the technology's benefits while minimizing associated risks. The proposed amendments aim to provide clarity and certainty around the legal standing of smart contracts and ensure their compatibility with existing regulatory frameworks. By doing so, we hope to facilitate the broader adoption of smart contracts and unlock their full potential as a transformative tool for manufacturing companies and individuals alike. In Section 3, we introduce our proposed methodology, followed by its implementation in Section 4. Finally, in Section 5, we conclude our research by highlighting the key findings and suggesting relevant managerial implications for companies considering implementing smart contracts.

## 2. LITERATURE REVIEW
### 2.1. Blockchain and smart contracts

What we call "smart contracts" are computer programs that contain pre-written instructions that can be followed without human intervention. Because to this flaw, a computer is unable to interpret the contract, which, in many instances, reduces the contract's intelligence (Kaulartz & Heckmann, 2016). Nik originally suggested the idea of smart contracts in the middle of the 19th century. There have been several forms of "smart contracts" for decades (O'Shields, 2017; Fauziah et al., 2020), such as the transaction processing systems used by banks to manage their daily payments and receipts. However, with the advent of Bitcoin and its underlying technology, Blockchain, this idea has been placed in new contexts with expanded powers. A platform can better leverage smart contracts thanks to the security and accuracy provided by blockchain technology (Liu & Liu, 2019; Lauslahti et al., 2017). When default contract terms are established, a smart contract will carry them out. When the terms of the agreement are met, the parties to the contract implement the operational and executive action plan outlined in the "smart contract" that they have signed using cryptographic security and have expanded in the distributed ledger, also known as the blockchain. For instance, after a service or product has been satisfactorily delivered, the corresponding payment can be processed by the smart contract via the decentralized platform (Corrales et al., 2019; Zhang et al., 2021). In the event that payment is not received, the smart contract will initiate the process of recovering the items or stopping the delivery of services. Financial instrument trading, syndicated loan transactions, and securities payments are only some of the many possible uses for this technology (Hewa et al., 2021; Nugent et al., 2016). The digital currency Bitcoin is encrypted and included in the blockchain technology. In 2008, a mysterious person using the alias Satoshi Nakamoto introduced the world to Bitcoin. Notwithstanding the decline in Bitcoin's popularity as a payment method, blockchain technology has recently received more attention, especially in the banking and finance sector (Alipour et al., 2021; Khorsandi and Bayat, 2022; Murray & Anisi, 2019; Griggs et al., 2018; Khorsandi and Khorsandi, 2022). All completed bitcoin transactions are recorded in the blockchain, which functions as a public ledger. Using cryptographic techniques and a substantial amount of computing power, a network of computers verifies each transaction or block before adding it to chains of all previous transactions. The distributed ledger, or blockchain, is completely public and accessible to anyone. The addresses displayed are not always those of the people who use such addresses, and the system is intended to protect users' anonymity (Destefanis, et al., 2018; Peters & Panayi, 2016; De Giovanni, 2020; Mehregan et al., 2023). This header should be considered static and unmodifiable. Many financial organizations and banks around the world are interested in this technology because of its security, longevity, and immutability. Blockchain's safety features include encryption and the usage of both public and private keys. Each user in a transaction is assigned a unique public address and a secret, mathematically-derived security access key (Papadodimas et al., 2018; Nayak et al., 2018). If these conditions are met, the transaction is announced to the rest of the blockchain's users for verification and recording. While "proof of work" is the mechanism Bitcoin uses for security, there are alternative ways to verify that transactions are legitimate and cannot be duplicated. One of the planned innovations for blockchain is the ability to track transactions without a third party (Pan et al., 2018). Possibility of transferring ownership of an asset directly from one owner to another, eliminating the requirement for a third party to act as a trusted intermediary. The goal of integrating smart contracts with blockchain technology is to facilitate the smooth completion of financial transactions on this network (Almasoud et al., 2020; Uriarte et al., 2018). Smart contracts are unique in certain ways, but they are not revolutionary in others. Agreements between parties with legal contracting power should be clear and unambiguous. Moreover, banking organizations have employed computerized automated systems for decades (Oliva et al., 2020; Larijani and Dehghani, 2023) to process transactions without the need for human participation.

## 2.2. Possibilities and constraints in the technical and business sectors

Blockchain, or distributed ledger technology, and smart contracts are complementary technologies (in the legal system). Smart contracts can improve global payments, syndicate lending, collateral management, proxy voting, securities issuance, and regulatory and compliance activities, according to the World Economic Forum (Kemmoe et al., 2020; Rozario & Thomas, 2019).

For instance, smart contracts might be used to create a lending syndicate and have that syndicate's financial and service needs met automatically. The central banks are attempting to launch blockchain-based digital currencies. Collateral submitted for a transaction can be tracked and evaluated and settled more quickly and easily with the help of smart contracts (Shojaei et al., 2020; Cohn et al., 2016).

ISDA master contracts, credit support attachments, and endorsements are all examples of derivative papers that the British bank Barclays has attempted to transform into automated smart contracts (Bodó et al., 2018; Abdellatif & Brousmiche, 2018; Cuocci et al., 2023). In the Barclays model, the core contracts are stored on a centralized distributed platform, and other copies of the contracts are made available for download and usage by interested parties. To promote creativity and teamwork in the financial sector's blockchain development, Barclays is making available the underlying technology it has been using in its initiatives as open source. A smart contract prototype for stock exchanges, including after-sale services, was recently tested successfully by many large banks, including JP Morgan and Credit Suisse. It's been very similar to making a wire transfer or processing a business. BNP Paribas, a French bank, also studies legal contracts that are automatically triggered (Balcerzak et al., 2022; Howell & Potgieter, 2021; Dehghani and Larijani, 2023; Sasani et al., 2023; Sadeghi et al., 2022).

There is also talk of using smart contracts for everyday financial dealings. While conducting business with a company online, it is in the best interest of customers to have more leverage in negotiations of commercial terms. It's possible that in this scenario, both buyers and sellers could employ automated purchasing agents to facilitate online transactions. This could pave the way for a future in cyberspace where intelligent contracts bargain with one another (Khan et al., 2020). Automatic automobile payments and easy access to rented properties are two other examples of consumer uses for smart contracts (Lin et al., 2022). Smart contracts are expected to improve the efficiency and accuracy of corporate transactions, the effectiveness of operations, and the speed and low cost of contract execution. More than a billion dollars will be spent on blockchain projects in 2017 by financial institutions. This makes blockchain one of the fastest growing enterprise software sectors. Despite $1.4 billion being invested over the past three years (Lin et al., 2022; Małkowska et al., 2021), this is the case. There is widespread anticipation that blockchain technology will debut in 2017. But, experts have warned that many of the use cases for blockchain and smart contracts are overly complicated and could end up costing a lot of money. At first, blockchain applications are likely to be implemented within the United States and will focus on the movement of data (information) rather than monetary payments (and then the technology will go global for large-scale electronic transactions for financial payments). Be successful in passing through). The full potential of blockchain technology is not likely to be realized in its current form of use until the financial services industry as a whole adopts common platforms or at least compatible operating systems (Mulligan, 2021). Both the cost-benefit analysis and the potential scope of this technology's adoption in the future remain unclear at this time. Due to the complexity of integrating the technology into the security and trust needs of financial institutions under stringent oversight (Fotiou et al., 2018), some observers have argued that opponents of "Blockchain technology" may have reached a new high. Smart contracts have many potential advantages, but they also carry some serious risks. Cybersecurity is a major concern with respect to smart contracts (Molina-Jimenez et al., 2018; Allam, 2018).

Can these predetermined agreements be hacked and redirected to pursue erroneous aims? What's more, can they genuinely replace conventional paper contracts without a way to modify and enforce them? In this setting, though, the latter is unfortunately ineffective (well as a substitute). An internal hacker in July 2016 exploited coding flaws in a so-called decentralized autonomous (non-profit) organization (DAO) to siphon off $50 million (Sharma et al., 2020; Lone & Naaz, 2021). Using the Ethereum platform for the development and expansion of smart contracts, the DAO was an investment fund created to run automatically, without administration or a board of directors. It's obvious that the hacker was an

insider rather than a random passer-by. Once the security flaw was discovered and patched, the stolen monies were returned. But, some DAO users became upset following the next code modification because they felt it watered down the integrity of blockchain and smart contracts. Even the most sophisticated contracts, as one observer pointed out, are not immune to the possibility of human error (Hasan & Salah, 2019; Palma et al., 2019).

**2.3. Problems with the Law, Number Three In Contract Rights**

The validity of smart contracts as legal agreements is an initial concern. It is possible to legally rely on a statement, promise, or contract. There are a number of requirements that the law imposes on a contract before it may be considered enforceable (Hu et al., 2018). Two parties, the parties' legal power to enter into the contract, mutual consent, and guarantors are required. Conditions essential to the performance of the contract include the absence of mistakes, manipulation, unwillingness, compulsion, and inconsistencies with universal norms. The usual rule is that a contract can be either verbal or written (Triana Casallas et al., 2020; Uriarte et al., 2021). Contracts of significance must be put in writing, and in practice, the vast majority of business contracts are put in writing, either in the form of a traditional written document or electronically, through the use of electronic terminology (Xuan et al., 2020; Daniel & Guida, 2019). A smart contract must adhere to the same standards as any other legally binding agreement in order to be enforced. The smart contract parties' agreement on certain terms may be one area to examine. Stakeholders can only be satisfied as a whole if their needs are communicated and met through performance reporting. Contracts can be concluded verbally or in writing, depending on the nature of the agreement between the parties, but as has been discussed, some agreements simply must be in physical form. Traditionally, the ideas of demand and acceptance of the parties in the contract are depending on the mutual satisfaction of the parties (Leka et al., 2019; Goldenfein & Leiter, 2018; Lamb, 2018).

Recent cases have explored the idea of contract creation in the digital age, but these studies have always relied on tried-and-true legal principles like the parties' informed consent. There are two common types of online contracts, "clicking" and "browsing," (Molina-Jimenez et al., 2019) each with its own set of characteristics. Unlike a browser-based agreement, which requires the user to affirmatively interact with a checkbox before agreeing to its terms, a contract that is clicked accepts its terms automatically. In most cases, showing "actual knowledge" of the contract's provisions is required to win in court. The user must be placed in a difficult position to become aware of the contract's conditions without having actual awareness of them. This normally necessitates the disclosure of the conditions and the drawing of the user's careful attention to the fact that their ongoing use of the website constitutes their obligation to the stated conditions. The United States Court of Appeals for the Ninth Circuit recently ruled (Khan et al., 2021; Hasan & Salah, 2018) that an outward show of agreement is insufficient to establish a binding contract.

Conclusions drawn from these examples show that in order for a smart contract to be legally binding, it must both explicitly communicate the conditions to the parties to the contract and have a clear background of mutual approval of the terms by the contract parties, such as the "agree" button being clicked. Agreements that meet these requirements have a better chance of being upheld in court. The court ruled in a recent instance that consent can be reached provided the following conditions are met (Novikov et al., 2018; Hamilton, 2020):

There must be: (1) a prominent notice outlining the conditions of use for online transactions; (2) a prominent notice that, in addition to the transaction, binds the contracting party to the terms contained in the contract; and (3) the existence of the user's explicit agreement with the contractual conditions when creating an account. In commercial law, even legally binding contracts can be invalidated on the basis of public policy considerations like uncertainty. The immutability of the blockchain could be at odds with the practice of reviewing and not enforcing a smart contract after its conclusion (Kim & Laskowski, 2017; Choudhury et al., 2018). Smart contracts' immutability makes the resulting transaction record legally binding. Public policy considerations are not the only reason a smart contract transaction might be evaluated by a court or the parties to the agreement after it has taken place. It's hard to imagine that major banks, regulators, or government officials would use technology that they can't tweak as needed. According to one expert, "smart contracts maintain the best elements of regular contracts, including the flexibility to negotiate if necessary in the future (Choudhury et al., 2018)."

Electronic signatures and contracts completed entirely based on electronic form are generally enforceable under federal law on electronic signatures in worldwide and national trade. In addition to requiring that electronic contracts be preserved in a readable, saveable, and retrievable manner, the aforementioned law also specifies that certain additional conditions must be met in electronic contracts, such as consumer notices in certain instances. Last but not least, the expansion and development of alternative electronic signature and registration regulations, such as the Electronic Transactions Standardization Law (Wright & Serguieva, 2017; Hoorfar et al., 2023; Moshtaghi

Largani & Lee, 2023), have been made possible by the electronic signature legislation.

When it comes to preparing state e-commerce laws and providing consistent rules for e-commerce transactions, Utah was the first to take a complete approach. Electronic signatures and documents not covered by the Uniform Commerce Act's Articles 2 and 2 are designed to be governed by Utah law (a). Not only that, but Utah law only applies if both parties agree to do business online. The Utah law was approved, according to the approach of the courts in using existing legal principles to conduct electronic transactions, not to create a completely new system of legal principles for electronic markets, but rather to guarantee that the terms of electronic contracts are as enforceable as those of non-digital ones. To be Utah was developed to supply guidelines for extending traditional legal principles to online dealings (Schrepel, 2019; Hunhevicz et al., 2022).

To comply with the Utah Act's regulations for electronic signatures, every document attributing an activity to a person must incorporate security processes to validate the legitimacy of the signature or the consent of the parties to the transaction (Section 9). The term of "note" in Article 3 of the UCC is expanded to include "transferable recordings" under Utah law, however electronic notes must be kept as the sole proof of the responsibilities and rights included in the note. In Utah's law, Section 4, contracts made with computer programs or other electronic agents are treated as valid. Particularly helpful for negotiators who have begun to adopt and employ artificial intelligence and robotics technology in the course of negotiations (Temte, 2019; Deebak & Fadi, 2021; Roudini et al., 2020; Gavidia et al., 2023; Kalantari et al., 2022; Abbasi et al., 2023; Dehghan and Naghibi Iravani, 2022; Dehghani and Larijani, 2023; Amiri et al., 2023; Malmir et al., 2023) are these provisions of the law.

It appears clear, based on the application of legal principles related to electronic transactions, that smart contracts do not require a distinct body of new laws or regulations; rather, existing legal principles will be adapted and, perhaps, modified through regulation or the judiciary, to precisely respond to the legal requirements of smart contracts and the corresponding technological advances. There are new technologies on the horizon, albeit there will probably be a significant lag time between widespread use of these innovations and any necessary legal changes. Legal criteria relating to the conclusion of the contract, such as showing the existence of mutual consent of the parties regarding the terms contained in the contract, must be met for smart contracts to be declared valid. To do so, a button can be provided with a declaration of agreement with the contractual conditions and a link to the terms written in plain English. The terms of the contract must be stored securely, so that they cannot be altered without the consent of all parties involved in the transaction. Computer-generated contracts. Also, they need to follow federal and state regulations concerning electronic transactions, such as Utah's E-SIGN Act (Rikken et al., 2019; Roszko-Wójtowicz & Grzelak, 2021; Rowland, 2022; Rozas et al., 2021).

### 2.4. Evidence and implementation concerns

To have the same legal force as a standard contract, smart contracts need to be in accordance with applicable federal and state regulations. In particular, difficulties with proof, the enforceability of defense-depriving conditions, jurisdiction, and the choice of governing law (Rymarczyk, 2020; Shan et al., 2021) can delay the effectiveness of such contracts. Given that the whole point of a smart contract is that it executes itself and does away with the need for human intervention, these implementation issues have the potential to lessen the benefits of smart contracts. Although smart contracts have the advantage of self-execution and electronic execution of contract requirements, human participation is still required to resolve legal disputes (Sifah et al., 2020). To resolve conflicts arising from electronic contracts, it has been calculated and implemented. Given that a smart contract is a piece of code, it is possible that there will be unique challenges in proving a claim under the contract. As it is highly improbable that the court will have the expertise to review the code directly, the code must be written in natural language for evaluation as part of the litigation. Predictable resolution of this issue is possible through the creation and maintenance of a code translation into plain language that can be updated whenever the smart contract terms are modified. The creators of this tech will find this a simple assignment. Reason being, they will help the parties reach an understanding by providing a more conversational take on the formal language of the contract. In addition, the security mechanisms must be proven to be adequate to preserve the code in the agreed state, which requires more knowledge and experience in smart contract technology (Singh & Chopra, 2020; Sinha & Roy Chowdhury, 2021).

The settlement of external disputes of smart contracts creates concerns with the enforceability of defense-depriving conditions, in addition to the evidentiary issues. These concerns with smart contracts may be settled by applying preexisting legal standards. Estimates, for instance, are not allowed to deviate more than is shown in conventional contracts after the elimination of limitations (Sinha & Roy Chowdhury, 2021; Sotoudehnia, 2021). This could make it more difficult to keep track of who gave what consent and when, which could lead to technological difficulties in completing transactions. It is highly doubtful that the current political and regulatory climate will result in the adoption of new legal standards for smart contracts, therefore the same standards that apply to paper contracts will likely be applied to smart contracts involving consumers and businesses (Sun et al., 2016; Viano et al., 2022). Last but not least, the execution of smart contracts on distributed ledgers like blockchain raises certain legal questions. When disagreements arise, for instance, what part does the letterhead that was sent out have in mediating the situation? As a blockchain-based system is anonymous, determining who was involved in a transaction is a grey area that may lead to more complications in the event of a dispute. In addition, the blockchain platform operator must be traceable and must have the financial wherewithal to be a party to the contract in the event of a disagreement. Alternately, the operator's or a third party's identification can be used to provide an appropriate forum for external conflict resolution (Wallace & Lăzăroiu, 2021; Watson, 2022).

When developing the platform, the platform operator may include (particular) legal provisions in the terms of service for the platform and all associated smart contracts. The legislation that will apply to any disputes that may arise out of a smart contract and the location where those disputes will be resolved must be made explicitly apparent. Participants can enter into traditional contracts at the time of constructing the platform by consenting to the legal provisions of the parties, including dispute resolution, governing legislation, and jurisdiction. Several of the proof-of-claims problems outlined above, including the "venue" of the platform and transactions, will play a role in determining the governing law and competent jurisdiction in the absence of an express agreement. Larger, more complicated transactions are less likely to benefit from post-contractual resolution of such matters (owing to their high sensitivity), (Wong et al., 2022; Yakymova et al., 2022). Four and Two-Half Difficulties in Countering Financial Crime.

Compliance with anti-terrorism and money-laundering regulations is another area where smart contracts may bring unique difficulties. Laws like these typically make it illegal to report "suspect conduct" to authorities or transfer money to regulated persons, as well as requiring anyone involved in financial transactions to know and authenticate the identities of traders. Users of smart contracts may need to implement controls to ensure compliance with these regulations, such as authenticating users and preventing fraudulent transfers and purchases by integrating with other Systems to automatically update lists of forbidden transactions, (Zachariadis et al., 2019; Chatterjee et al., 2019).

### 2.5. Legal ethics

A wide range of professional obligations, including the commission of unlawful conduct, could be addressed in smart contracts. Lawyers are forbidden from engaging in unlicensed practice of law under professional standards. If you're not a lawyer and you want to split your legal expenses with a non-lawyer or form a business partnership with one, you could face criminal charges in various US jurisdictions. In addition to being against the rules of legal ethics, breaking the law is illegal in the majority of states. In the legal field, one engages in a variety of activities, such as the creation of legal documents, the rendering of legal opinions, and the provision of legal services or guidance (Ramachandran & Kantarcioglu, 2017, Sekhar et al., 2019). While lawyers are permitted by law to use the services of others to aid them in their practice, they are nonetheless held to the highest standards of professional responsibility for all work product. The application of smart contract technology, in which the contract exists as computer code, which is then treated as a legal contract, may be complicated by these rules (Drummer & Neumann, 2020; Sultana et al., 2020). A lawyer working on such a project will collaborate with computer scientists to verify that the code properly captures legal words and other nuances of natural language. This activity, such as creating legal terms and conditions, can be very useful in the competitive online electronic market. Large-scale agreements may make use of smart contracts since they are able to deal with complex issues involving multiple parties with specialized knowledge, such major banks and other financial traders. A consultant will function as a representative for both parties to the transaction, and this consultant will need to be familiar with smart contract technology. Finally, the lawyer must check that the provisions of the contract are fixed in computer code and will remain secure and unchanged for the duration of the agreement (Ariyarathna et al., 2019; Kongmanee et al., 2019).

### 3. RESEARCH METHODOLOGY

The research methodology for this study involved the use of two techniques: The Delphi method and Stepwise Weight Assessment Ratio Analysis (SWARA). The Delphi method was used to identify critical success factors for applying blockchain transactions in a manufacturing company. This technique involves a group of experts answering questionnaires, with the responses being analyzed and fed back to the participants. The process is repeated until a consensus is reached. Following the Delphi method, SWARA was used to determine the most influential factors among the identified success factors. SWARA is a multi-criteria decision-making technique that ranks alternatives based on their weight and importance. The process involves several steps, including defining criteria, weighting them, normalizing weights, determining weighted values, and calculating the final score for each alternative. The proposed methodology was then implemented, and data were collected through a structured questionnaire distributed to a panel of industry experts. The survey was designed to elicit responses on various factors affecting the successful implementation of smart contracts in a manufacturing company. The results of the survey were analyzed using the Delphi and SWARA methods to identify the most influential factors for the successful application of blockchain transactions as smart contracts in a manufacturing company. The final outcome of the research was a list of critical success factors, ranked according to their importance, which can contribute to the development of a legal framework for smart contracts in a manufacturing company. By utilizing these critical success factors, businesses can increase the likelihood of successfully implementing blockchain transactions as smart contracts in the manufacturing industry.

### 3.1. Stepwise Weight Assessment Ratio Analysis (SWARA) method

The SWARA approach uses a comparison to establish how important various factors are, while calculating SWARA, methods are prioritized according to their relative impact.

*Step 1: A point value is assigned to each criterion. Rankings are shown as Relative Value Averages or $S_j$.*

*Step 2: Coefficient $k_j$ can be computed as follows:*

$$k_j = \begin{cases} 1, & j = 1 \\ s_j + 1, & j > 1 \end{cases} \quad (1)$$

*Step 3: importance indicators of $q_j$ calculate as follows:*

$$q_j = \begin{cases} 1, & j = 1 \\ \frac{k_{j-1}}{k_j}, & j > 1 \end{cases} \quad (2)$$

*Step 4: Following is the formula used to determine criterion weights::*

$$w_j = \frac{q_j}{\sum_{k=1}^{n} q_k} \quad (3)$$

The relative weight of criterion j will be illustrated as $w_j$

Hashemkhani Zolfani develops an enlargement of the SWARA approach (2018). They determine $t_{jk}$ values using this novel technique. Expert interview data and the average of ($\bar{t_j}$) are used to calculate the following equation.

$$\bar{t_j} = \frac{\sum_{k=1}^{r} t_{jk}}{r} \quad (4)$$

$t_{jk}$ reflect the order in which the attributes j and k were answered and r show the total number of responses. Following this, the equation below can be used to determine the relative importance of the traits revealed by q j.

$$q_j = \frac{\bar{t_j}}{\sum_{j=1}^{n} t_j} \quad (5)$$

The total weights of $q_j$ are an equal one. The variation of values are obtained by

$$\sigma^2 = \frac{1}{r-1} \sum_{k=1}^{r} (t_{jk} - \bar{t_j})^2 \quad (6)$$

$$\beta_j = \frac{\sigma}{\bar{t_j}} \quad (7)$$

Then, we may use an equation to derive W's value... W indicates the consistency between the opinions of specialists and hence symbolizes the credibility of the data.

$$W = \frac{12s}{r^2(n^3-n) - r\sum_{k=1}^{r} T_k}, \quad W \in [0,1], \quad (8)$$

S is the sum of the squared deviations of the attributes and n is the number of experts; T k is the index of the number of times k ranks in r. A measure of attribute ranking dispersion is calculated:

$$S = \sum_{j=1}^{n} \left[ \sum_{k=1}^{r} t_{jk} - \frac{1}{n} \sum_{j=1}^{n} \sum_{k=1}^{r} t_{kj} \right]^2 \quad (9)$$

### 3.2. Delphi method

The Delphi technique surveys the views of unnamed experts and attaches them to a strict timetable in the form of written, argumentative, and reactive arrangements. With the use of in-depth observation and a lack of direct interaction, this technique claims to speed up the group's conclusion-drawing process. The strategy for systematic for a range of choices on a specific theme flowing intended sequential intervals, a buffet with snippets of facts and critique of opinions that flow from earlier replies (Delbecq et al. 1975; Osborne et al. 2003). Whilst the Delphi technique is regularly scheduled to help offer expert thoughts, it is rumored to be time-consuming, costly, and still leaves room for some vagueness and nebulousness in the responses of the specialists polled (Chang et al. 2000). The Delphi technique, which was endorsed by Murry and Hammons (1995), provides a recap of 10–30 expert estimates for the array.

The Delphi technique is being looked at as a possible means of expert-driven CSF customization (Brady, S. R. 2015). The CSFs will be used to design the questionnaire, and the experts' feedback will be collected using Likert-type scales (Strand et al. 2017). For instance, when experts using a 5-point Likert-scale, if the mean score is below 4, all but two of the ensuing CSFs are deemed unnecessary. It is recommended to use between 5 and 15 specialists for this survey technique, data miners' information consists in table 1:

Table1: Survey-based data mining information

| Expert | Education | Experience |
|---|---|---|
| Expert.1 | Ph.D. | 10 |
| Expert.2 | MASTER. | 11 |
| Expert.3 | Ph.D. | 14 |
| Expert.4 | Ph.D. | 22 |
| Expert.5 | MASTER | 9 |
| Expert.6 | MASTER | 12 |
| Expert.7 | Ph.D. | 8 |

## 3.3. Critical Successful Factors

Twenty-six cerebrospinal fluids (CSFs) were effectively collected in this investigation based on a previous one. Next, we used the Delphi technique to tailor these CSFs to our needs. The formula for determining individual CSFs is displayed in Table No. 2.

Table 2: Calculating individual CSFs

| No. | criteria | code |
|---|---|---|
| 1 | Manufacturing Plants | PF |
| 2 | Purpose of Quality Management | QMI |
| 3 | System Outcome Quality | QSO |
| 4 | Claims | CL |
| 5 | Enhancement of Quality | QI |
| 6 | Delivery | De |
| 7 | The Counter Argument | RC |
| 8 | Prompt shipping | OD |
| 9 | In charge: administration and structure | MO |
| 10 | Management of an Organization | OC |
| 11 | Strategic plans | BP |
| 12 | Conversations with Clients | CC |
| 13 | A Check From Inside | IA |
| 14 | Management of information | DA |
| 15 | Vision | CO |
| 16 | Stability in the bank account | FL |
| 17 | In a Healthy Way Environment | VIS |
| 18 | Mechanics of Coordination | FOP |
| 19 | Connections with government entities | HSE |
| 20 | Subcontractors The Guarantee of Quality | ENC |
| 21 | Turnover | RPA |
| 22 | Components for Building | SQA |
| 23 | Strategies for Subcontractors | TNO |
| 24 | The Proposal's Potential Social Effects | COR |
| 25 | Cost | SUS |
| 26 | Dependability Constant | SIP |
| 27 | adaptability | COS |

## 4. FINDING

The Delphi technique takes into account the input of unnamed subject matter experts and attaches their opinions to the predetermined timetable of written arguments and responses. As face-to-face discussions have disadvantages, the Delphi method posits that professionals draw conclusions through observation of alternative views. The foundation of this research strategy is the continuous and gradual incorporation of participant feedback. Every step along the way, the expert's fresh perspective will inform an attempt to refine the earlier findings. Delbecq et al. (1975) and Osborne et al. (2003) both support this. The Delphi technique is useful for incorporating professional suggestions into a project in a well-organized fashion; nevertheless, it is time-consuming, expensive, and fraught with ambiguity; and there are counterarguments to be found in the experts' opinions (Chang et al., 2000; Anbari et al., 2020a; Anbari et al., 2020b). The Delphi method, endorsed by Murry and Hammons (1995), synthesizes the range of 10–30 from the experts' estimates.

The method has been seen as a tool for tailoring CSFs in light of specialist feedback (Brady, S. R., 2015). In light of these CSFs, a questionnaire would be developed for experts to express their views on a predetermined Likert scale (Strand et al., 2017). By way of illustration, if the experts' average rating on a 5-point Likert scale is below 4, the ensuing CSFs will be discarded. Specifically, the research indicates

Table 3: The findings of four CSFs

| Code | Expert 1 | Expert 2 | Expert 3 | Expert 4 | Expert 5 | Expert 6 | Expert 7 | Average | Accept/Reject |
|---|---|---|---|---|---|---|---|---|---|
| PF | 4 | 3 | 4 | 4 | 1 | 3 | 5 | 3.428571 | Accept |
| QMI | 5 | 4 | 3 | 5 | 4 | 3 | 2 | 3.714286 | Accept |
| QSO | 1 | 4 | 4 | 1 | 1 | 3 | 1 | 2.142857 | Accept |
| CL | 1 | 5 | 2 | 3 | 3 | 5 | 5 | 3.428571 | Accept |
| QI | 4 | 1 | 5 | 2 | 1 | 5 | 3 | 3 | Accept |
| De | 5 | 5 | 3 | 4 | 5 | 3 | 3 | 4 | Accept |
| RC | 1 | 4 | 3 | 4 | 5 | 1 | 3 | 3 | Accept |
| OD | 3 | 2 | 3 | 3 | 5 | 5 | 1 | 3.142857 | Accept |
| MO | 1 | 3 | 1 | 2 | 2 | 4 | 1 | 2 | Reject |
| OC | 5 | 5 | 5 | 5 | 5 | 3 | 3 | 4.428571 | Reject |
| BP | 2 | 5 | 4 | 1 | 1 | 3 | 1 | 2.428571 | Reject |
| CC | 3 | 2 | 2 | 3 | 2 | 4 | 1 | 2.428571 | Accept |
| IA | 4 | 3 | 4 | 2 | 5 | 5 | 2 | 3.571429 | Accept |
| DA | 3 | 3 | 1 | 5 | 5 | 4 | 4 | 3.571429 | Reject |
| CO | 5 | 4 | 2 | 3 | 4 | 1 | 1 | 2.857143 | Accept |
| FL | 1 | 2 | 3 | 3 | 2 | 4 | 3 | 2.571429 | Accept |
| VIS | 5 | 3 | 3 | 2 | 1 | 4 | 5 | 3.285714 | Accept |
| FOP | 2 | 4 | 3 | 2 | 4 | 2 | 1 | 2.571429 | Accept |
| HSE | 2 | 3 | 1 | 4 | 3 | 2 | 1 | 2.285714 | Accept |
| ENC | 4 | 5 | 1 | 2 | 2 | 1 | 3 | 2.571429 | Accept |
| RPA | 1 | 2 | 2 | 3 | 4 | 1 | 5 | 2.571429 | Accept |
| SQA | 5 | 1 | 1 | 3 | 4 | 3 | 4 | 3 | Accept |
| TNO | 1 | 5 | 2 | 4 | 1 | 5 | 3 | 3 | Accept |
| COR | 5 | 4 | 1 | 1 | 4 | 2 | 1 | 2.571429 | Accept |
| SUS | 4 | 5 | 4 | 1 | 4 | 5 | 1 | 3.428571 | Accept |
| SIP | 3 | 3 | 2 | 5 | 1 | 1 | 2 | 2.428571 | Accept |
| COS | 2 | 2 | 1 | 5 | 3 | 5 | 1 | 2.714286 | Accept |

that between five and fifteen technique experts are required.

According to the findings, four of the CSFs (new technology, modularization of manufacturing, product definition, product competency, and product presentation within the particular period) were deemed unnecessary by the experts, in table 3:

After screening CSFs, first primary weights are obtained based on the SWARA method, demonstrated in Table 4.

turnover, the counter argument, vision, components for building, and system outcome quality are the most influential factors for successful implementation. Conversely, connections with government entities and subcontractors, and the guarantee of quality have the least influence. These findings can contribute to the development of a legal framework for smart contracts in a manufacturing company. Overall, further research is needed to address the challenges associated with the enforcement of smart contracts and to establish a regulatory framework that can provide clarity and

Table 4: SWARA-based weights

| Attributes | Comparative importance of average value ($s_j$) | Coefficient $K_j = S_j + 1$ | Weight $q_j = \dfrac{W_j}{\sum W_j}$ |
|---|---|---|---|
| Manufacturing Plants | 0.27142857 | 1.271429 | 0.044533 |
| Purpose of Quality Management | 0.34285714 | 1.342857 | 0.047035 |
| System Outcome Quality | 0.35714286 | 1.357143 | 0.047536 |
| Claims | 0.28571429 | 1.285714 | 0.045034 |
| Enhancement of Quality | 0.24285714 | 1.242857 | 0.043533 |
| Delivery | 0.25714286 | 1.257143 | 0.044033 |
| The Counter Argument | 0.41428571 | 1.414286 | 0.049537 |
| Prompt shipping | 0.3 | 1.3 | 0.045534 |
| In charge: administration and structure | 0.32857143 | 1.328571 | 0.046535 |
| Management of an Organization | 0.32857143 | 1.328571 | 0.046535 |
| Strategic plans | 0.2 | 1.2 | 0.042032 |
| Conversations with Clients | 0.3 | 1.3 | 0.045534 |
| A Check From Inside | 0.34285714 | 1.342857 | 0.047035 |
| Management of information | 0.31428571 | 1.314286 | 0.046035 |
| Vision | 0.37142857 | 1.371429 | 0.048036 |
| Stability in the bank account | 0.35714286 | 1.357143 | 0.047536 |
| In a Healthy Way Environment | 0.22857143 | 1.228571 | 0.043032 |
| Mechanics of Coordination | 0.3 | 1.3 | 0.045534 |
| Connections with government entities | 0.24285714 | 1.242857 | 0.043533 |
| Subcontractors The Guarantee of Quality | 0.24285714 | 1.242857 | 0.043533 |
| Turnover | 0.2 | 1.2 | 0.054 |
| Components for Building | 0.32857143 | 1.328571 | 0.048 |

## 5. CONCLUSION

In conclusion, the development of smart contracts has been made possible by the rise of digital currency and blockchain technology. These contracts offer several benefits, including the ability to execute automatically, provide public visibility of contract provisions, prevent financial crimes, and avoid contract abuses. However, enforcing smart contracts presents unique challenges that require human intervention, such as evidentiary issues, enforceability of waivers of defenses, and jurisdictional and choice-of-law considerations. The lack of standardized regulations poses a significant challenge for their widespread adoption, with countries relying on customary law to legitimize their use. The Delphi method and SWARA were used in this study to identify critical success factors for applying blockchain transactions as smart contracts in a manufacturing company. Results show that security for users. With continued advancements in technology and regulation, smart contracts have the potential to transform various industries, including manufacturing.

**REFERENCES**

Abbasi, H., Orouskhani, M., Asgari, S., & Zadeh, S. S. (2023). Automatic Brain Ischemic Stroke Segmentation with Deep Learning: A Review. *Neuroscience Informatics*, 100145. https://doi.org/10.1016/j.neuri.2023.100145

Abbasi, M., Manshaei, M. H., Rahman, M. A., Akkaya, K., & Jadliwala, M. (2022). On Algorand Transaction Fees: Challenges and Mechanism Design. In ICC 2022-IEEE International Conference on Communications (pp. 5403-5408). IEEE. https://doi.org/10.1109/ICC45855.2022.9838795

Abdellatif, T. and Brousmiche, K.L., 2018. Formal verification of smart contracts based on users and blockchain behaviors models. In *2018 9th IFIP International Conference on New Technologies, Mobility and Security (NTMS)* (pp. 1-5). IEEE.

Alipour, P., & Charandabi, S. E. 2023. Analyzing the Interaction between Tweet Sentiments and Price Volatility of Cryptocurrencies. *European Journal of Business and Management Research*, 8(2), 211-215.

Alipour, P., Foroush Bastani, A., & Mansourfar, G. 2021. Investigating the Performance of Portfolio Insurance Strategies under a Regime Switching Markov Model in Tehran Stock Exchange. *Financial Research Journal*, 23(2), 269-293.

Allam, Z., 2018. On smart contracts and organisational performance: A review of smart contracts through the blockchain technology. *Review of Economic and Business Studies*, *11*(2), pp.137-156.

Almasoud, A.S., Hussain, F.K. and Hussain, O.K., 2020. Smart contracts for blockchain-based reputation systems: A systematic literature review. *Journal of Network and Computer Applications*, *170*, p.102814.

Amiri, M. K., Zaferani, S. P. G., Emami, M. R. S., Zahmatkesh, S., Pourhanasa, R., Namaghi, S. S., ... & Hajiaghaei-Keshteli, M. (2023). Multi-objective optimization of thermophysical properties GO powders-DW/EG Nf by RSM, NSGA-II, ANN, MLP and ML. *Energy*, 128176. https://doi.org/10.1016/j.energy.2023.128176

Anbari, M., Arıkan Öztürk, E. B. R. U., & Ateş, H. (2020a). Evaluation of sustainable transport strategies for Tehran with thetheir urbanization rate criterion based on the fuzzy ahp method. *Journal of Xi'xxan University of Architecture Technology*, 12(7), pp. 867-881.

Anbari, M., Arıkan Öztürk, E. B. R. U., & Ateş, H. A. K. A. N. (2020b). Urban Design And Upgrading Traffic And Urban Street Safety From The Perspective Of Urban Users (Case Study: Tunalı Hilmi-Ankara Residential Commercial Street). *The Journal of International Social Research*, 13(69), pp. 506-515.

Ante, L., 2021. Smart contracts on the blockchain–A bibliometric analysis and review. *Telematics and Informatics*, *57*, p.101519.

Ariyarathna, T., Harankahadeniya, P., Isthikar, S., Pathirana, N., Bandara, H.D. and Madanayake, A., 2019, April. Dynamic spectrum access via smart contracts on blockchain. In *2019 IEEE Wireless Communications and Networking Conference (WCNC)* (pp. 1-6). IEEE.

Balcerzak, A.P., Nica, E., Rogalska, E., Poliak, M., Klieštik, T. and Sabie, O.M., 2022. Blockchain technology and smart contracts in decentralized governance systems. *Administrative Sciences*, *12*(3), p.96.

Bodó, B., Gervais, D. and Quintais, J.P., 2018. Blockchain and smart contracts: the missing link in copyright licensing?. *International Journal of Law and Information Technology*, *26*(4), pp.311-336.

Chatterjee, K., Goharshady, A.K. and Pourdamghani, A., 2019, May. Probabilistic smart contracts: Secure randomness on the blockchain. In *2019 IEEE International Conference on Blockchain and Cryptocurrency (ICBC)* (pp. 403-412). IEEE.

Choudhury, O., Sarker, H., Rudolph, N., Foreman, M., Fay, N., Dhuliawala, M., Sylla, I., Fairoza, N. and Das, A.K., 2018. Enforcing human subject regulations using blockchain and smart contracts. *Blockchain in healthcare Today*.

Cohn, A., West, T. and Parker, C., 2016. Smart after all: Blockchain, smart contracts, parametric insurance, and smart energy grids. *Geo. L. Tech. Rev.*, *1*, p.273.

Cong, L.W. and He, Z., 2019. Blockchain disruption and smart contracts. *The Review of Financial Studies*, *32*(5), pp.1754-1797.

Corrales, M., Fenwick, M. and Haapio, H. eds., 2019. *Legal Tech, Smart Contracts and Blockchain*. Singapore: Springer.

Cuocci, S., Fattahi Marnani, P., Khan, I., & Roberts, S. (2023). A Meta-Synthesis of Technology-Supported Peer Feedback in ESL/EFL Writing Classes Research: A Replication of Chen's Study. Languages, 8(2), p. 114.

Daniel, F. and Guida, L., 2019. A service-oriented perspective on blockchain smart contracts. *IEEE Internet Computing*, *23*(1), pp.46-53.

De Giovanni, P., 2020. Blockchain and smart contracts in supply chain management: A game theoretic model. *International Journal of Production Economics*, *228*, p.107855.

Deebak, B.D. and Fadi, A.T., 2021. Privacy-preserving in smart contracts using blockchain and artificial intelligence for cyber risk measurements. *Journal of Information Security and Applications*, *58*, p.102749.

Dehghan, S., Naghibi Iravani, S. (2022). Comparison of seismic behavior factors for reinforced concrete (RC) special moment resisting frames (SMRFs) in Iran in low-, mid-, and high-rise buildings based on Iranian seismic standard 2800 and ASCE, *Journal of Economics and Administrative Sciences*, 5(S1), pp. 744-750.

Dehghani, F., & Larijani, A. (2023). A Machine Learning-Jaya Algorithm (Ml-Ijaya) Approach for Rapid Optimization Using High Performance Computing. Available at SSRN 4423338. https://dx.doi.org/10.2139/ssrn.4423338

Dehghani, F., & Larijani, A. (2023). An Algorithm for Predicting Stock Market's Index Based on MID Algorithm and Neural Network. Available at SSRN 4448033. https://ssrn.com/abstract=4448033

Destefanis, G., Marchesi, M., Ortu, M., Tonelli, R., Bracciali, A. and Hierons, R., 2018, March. Smart contracts vulnerabilities: a call for blockchain software engineering?. In *2018 International Workshop on Blockchain Oriented Software Engineering (IWBOSE)* (pp. 19-25). IEEE.

Drummer, D. and Neumann, D., 2020. Is code law? Current legal and technical adoption issues and remedies for blockchain-enabled smart contracts. *Journal of information technology*, *35*(4), pp.337-360.

Fauziah, Z., Latifah, H., Omar, X., Khoirunisa, A. and Millah, S., 2020. Application of blockchain technology in smart contracts: a systematic literature review. *Aptisi Transactions on Technopreneurship (ATT)*, *2*(2), pp.160-166.

Fotiou, N., Siris, V.A. and Polyzos, G.C., 2018. Interacting with the Internet of Things using smart contracts and blockchain technologies. In *Security, Privacy, and Anonymity in Computation, Communication, and Storage: 11th International Conference and Satellite Workshops, SpaCCS 2018, Melbourne, NSW, Australia, December 11-13, 2018, Proceedings 11* (pp. 443-452). Springer International Publishing.

Gatteschi, V., Lamberti, F., Demartini, C., Pranteda, C. and Santamaría, V., 2018. Blockchain and smart contracts for insurance: Is the technology mature enough?. *Future internet*, *10*(2), p.20.

Gavidia, J. C. R., Chinelatto, G. F., Basso, M., da Ponte Souza, J. P., Soltanmohammadi, R., Vidal, A. C., ... & Mohammadizadeh, S. (2023). Utilizing integrated artificial intelligence for characterizing mineralogy and facies in a pre-salt carbonate reservoir, Santos Basin, Brazil, using cores, wireline logs, and multi-mineral petrophysical evaluation. *Geoenergy Science and Engineering*, 231, p. 212303.

Goldenfein, J. and Leiter, A., 2018. Legal engineering on the blockchain:'Smart contracts' as legal conduct. *Law and Critique*, *29*, pp.141-149.

Griggs, K.N., Ossipova, O., Kohlios, C.P., Baccarini, A.N., Howson, E.A. and Hayajneh, T., 2018. Healthcare blockchain system using smart contracts for secure automated remote patient monitoring. *Journal of medical systems*, *42*, pp.1-7.

Hamilton, M., 2020. Blockchain distributed ledger technology: An introduction and focus on smart contracts. *Journal of Corporate Accounting & Finance*, *31*(2), pp.7-12.

Hasan, H.R. and Salah, K., 2018. Proof of delivery of digital assets using blockchain and smart contracts. *IEEE Access*, *6*, pp.65439-65448.

Hasan, H.R. and Salah, K., 2019. Combating deepfake videos using blockchain and smart contracts. *Ieee Access*, *7*, pp.41596-41606.

Hewa, T., Ylianttila, M. and Liyanage, M., 2021. Survey on blockchain based smart contracts: Applications, opportunities and challenges. *Journal of Network and Computer Applications*, *177*, p.102857.

Hewa, T.M., Hu, Y., Liyanage, M., Kanhare, S.S. and Ylianttila, M., 2021. Survey on blockchain-based smart contracts: Technical aspects and future research. *IEEE Access*, *9*, pp.87643-87662.

Hoorfar, H., Largani, S. M., Rahimi, R., & Bagheri, A. (2023). Minimizing Turns in Watchman Robot Navigation: Strategies and Solutions. arXiv preprint arXiv:2308.10090. https://doi.org/10.48550/arXiv.2308.10090

Howell, B.E. and Potgieter, P.H., 2021. Uncertainty and dispute resolution for blockchain and smart contract institutions. *Journal of Institutional Economics*, *17*(4), pp.545-559.

Hu, Y., Liyanage, M., Mansoor, A., Thilakarathna, K., Jourjon, G. and Seneviratne, A., 2018. Blockchain-based smart contracts-applications and challenges. *arXiv preprint arXiv:1810.04699*.

Hunhevicz, J.J., Motie, M. and Hall, D.M., 2022. Digital building twins and blockchain for performance-based (smart) contracts. *Automation in Construction*, *133*, p.103981.

Ibrahim, H., and Sasani, F. (2021). Study of The Interactions of Human Resource Performance and Electronic Human Resource Management in Small Companies, *Journal of Humanities Insights,* 5 (01), pp. 24-32.

Idelberger, F., Governatori, G., Riveret, R. and Sartor, G., 2016. Evaluation of logic-based smart contracts for blockchain systems. In *Rule Technologies. Research, Tools, and Applications: 10th International Symposium, RuleML 2016, Stony Brook, NY, USA, July 6-9, 2016. Proceedings 10* (pp. 167-183). Springer International Publishing.

Kalantari, R., Moqadam, R., Loghmani, N., Allahverdy, A., Shiran, M. B., & Zare-Sadeghi, A. (2022). Brain tumor segmentation using hierarchical combination of fuzzy logic and cellular automata. *Journal of Medical Signals and Sensors*, 12(3), 263. https://doi.org/10.4103%2Fjmss.jmss_128_21

Karami, T., Talebi, M. A., & Sabzevari, P. (2019). The impact of power sources and bureaucrats-orientation in predicting burnout (Case study: Employees of Shiraz Tax Administration). *Biannual Journal of Psychological Research in Management*, 5(1), pp. 48-77.

Kaulartz, M. and Heckmann, J., 2016. Smart Contracts–Anwendungen der Blockchain-

Technologie. *Computer und Recht*, *32*(9), pp.618-624.

Kemmoe, V.Y., Stone, W., Kim, J., Kim, D. and Son, J., 2020. Recent advances in smart contracts: A technical overview and state of the art. *IEEE Access*, *8*, pp.117782-117801.

Khan, S., Amin, M.B., Azar, A.T. and Aslam, S., 2021. Towards interoperable blockchains: A survey on the role of smart contracts in blockchain interoperability. *IEEE Access*, *9*, pp.116672-116691.

Khan, S.N., Loukil, F., Ghedira-Guegan, C., Benkhelifa, E. and Bani-Hani, A., 2021. Blockchain smart contracts: Applications, challenges, and future trends. *Peer-to-peer Networking and Applications*, *14*, pp.2901-2925.

Khan, Z., Abbasi, A.G. and Pervez, Z., 2020. Blockchain and edge computing–based architecture for participatory smart city applications. *Concurrency and Computation: Practice and Experience*, *32*(12), p.e5566.

Khorsandi, H., & Bayat, M. (2022). Prioritizing operational strategies of saman bank. *International Journal of Health Sciences*, 6(S7), pp. 1442–1453. https://doi.org/10.53730/ijhs.v6nS7.11548

Khorsandi, H., & Khorsandi, R. (2022). Ranking the effective factors in creative marketing in Iran Novin insurance. *Journal of Positive School Psychology*, 6(5), pp. 10009-10020.

Kim, H. and Laskowski, M., 2017, July. A perspective on blockchain smart contracts: Reducing uncertainty and complexity in value exchange. In *2017 26th International conference on computer communication and networks (ICCCN)* (pp. 1-6). IEEE.

Kongmanee, J., Kijsanayothin, P. and Hewett, R., 2019, November. Securing smart contracts in blockchain. In *2019 34th IEEE/ACM International Conference on Automated Software Engineering Workshop (ASEW)* (pp. 69-76). IEEE.

Kosba, A., Miller, A., Shi, E., Wen, Z. and Papamanthou, C., 2016, May. Hawk: The blockchain model of cryptography and privacy-preserving smart contracts. In *2016 IEEE symposium on security and privacy (SP)* (pp. 839-858). IEEE.

Lamb, K., 2018. Blockchain and Smart Contracts: What the AEC sector needs to know.

Larijani, A., & Dehghani, F. (2023). Stock Price Prediction Using the Combination of Firefly (FA) and Genetic Algorithms. Available at SSRN 4448024. https://ssrn.com/abstract=4448024

Laube, A., Sasani, F. (2020). Analysis of Risk Measurement in Financial Companies, *Journal of Humanities Insights*, 4 (01), pp. 27-28.

Lauslahti, K., Mattila, J. and Seppala, T., 2017. Smart contracts–How will blockchain technology affect contractual practices?. *Etla Reports*, (68).

Leka, E., Selimi, B. and Lamani, L., 2019, September. Systematic literature review of blockchain applications: Smart contracts. In *2019 International Conference on Information Technologies (InfoTech)* (pp. 1-3). IEEE.

Lin, R., Wang, L., Li, B., Lu, Y., Qi, Z. and Xie, L., 2022. Organizational governance in the smart era: The implications of blockchain. *Nankai Business Review International*, (ahead-of-print).

Liu, J. and Liu, Z., 2019. A survey on security verification of blockchain smart contracts. *IEEE Access*, *7*, pp.77894-77904.

Lone, A.H. and Naaz, R., 2021. Applicability of Blockchain smart contracts in securing Internet and IoT: A systematic literature review. *Computer Science Review*, *39*, p.100360.

Małkowska, A., Urbaniec, M. and Kosała, M., 2021. The impact of digital transformation on European countries: Insights from a comparative analysis. *Equilibrium. Quarterly Journal of Economics and Economic Policy*, *16*(2), pp.325-355.

Malmir, M., Momeni, H., & Ramezani, A. (2019). Controlling megawatt class WECS by ANFIS network trained with modified genetic algorithm. In 2019 27th Iranian Conference on Electrical Engineering (ICEE) (pp. 939-943). IEEE.

Mehregan, E., Sanaei, S., Manna, M., Bozorgkhou, H., & Heidari, S. (2023). The Role of SCM practices in Competitive Advantage and Firm Performance: A Mediating Role of Supply Chain Innovation and TQM. *Tehnički glasnik*, 17(4), pp. 516-523.

Molina-Jimenez, C., Sfyrakis, I., Solaiman, E., Ng, I., Wong, M.W., Chun, A. and Crowcroft, J., 2018, November. Implementation of smart contracts using hybrid architectures with on and off–blockchain components. In *2018 IEEE 8th International Symposium on Cloud and Service Computing (SC2)* (pp. 83-90). IEEE.

Molina-Jimenez, C., Solaiman, E., Sfyrakis, I., Ng, I. and Crowcroft, J., 2019. On and off-blockchain enforcement of smart contracts. In *Euro-Par 2018: Parallel Processing Workshops: Euro-Par 2018 International Workshops, Turin, Italy, August 27-28, 2018, Revised Selected Papers 24* (pp. 342-354). Springer International Publishing.

Moshtaghi Largani, S., & Lee, S. (2023). Efficient Sampling for Big Provenance. In Companion Proceedings of the ACM Web Conference 2023 (pp. 1508-1511).
https://doi.org/10.1145/3543873.3587556

Mulligan, K., 2021. Computationally networked urbanism and advanced sustainability analytics in internet of things-enabled smart city governance.

*Geopolitics, History, and International Relations*, *13*(2), pp.121-134.

Murray, Y. and Anisi, D.A., 2019, June. Survey of formal verification methods for smart contracts on blockchain. In *2019 10th IFIP International Conference on New Technologies, Mobility and Security (NTMS)* (pp. 1-6). IEEE.

Nayak, S., Narendra, N.C., Shukla, A. and Kempf, J., 2018, July. Saranyu: Using smart contracts and blockchain for cloud tenant management. In *2018 IEEE 11th International Conference on Cloud Computing (CLOUD)* (pp. 857-861). IEEE.

Niyafard S, Jalalian SS, Damirchi F., Jazayerifar S., Heidari S. (2023). Exploring the Impact of Information Technology on the Relationship between Management Skills, Risk Management, and Project Success in Construction Industries. *International Journal of Business Continuity and Risk Management*, 1-22. http://dx.doi.org/10.1504/IJBCRM.2024.10059509

Novikov, S.P., Kazakov, O.D., Kulagina, N.A. and Azarenko, N.Y., 2018, September. Blockchain and smart contracts in a decentralized health infrastructure. In *2018 IEEE International Conference" Quality Management, Transport and Information Security, Information Technologies"(IT&QM&IS)* (pp. 697-703). IEEE.

Nugent, T., Upton, D. and Cimpoesu, M., 2016. Improving data transparency in clinical trials using blockchain smart contracts. *F1000Research*, *5*.

Oliva, G.A., Hassan, A.E. and Jiang, Z.M., 2020. An exploratory study of smart contracts in the Ethereum blockchain platform. *Empirical Software Engineering*, *25*, pp.1864-1904.

O'Shields, R., 2017. Smart contracts: Legal agreements for the blockchain. *NC Banking Inst.*, *21*, p.177.

Palma, L.M., Vigil, M.A., Pereira, F.L. and Martina, J.E., 2019. Blockchain and smart contracts for higher education registry in Brazil. *International Journal of Network Management*, *29*(3), p.e2061.

Pan, J., Wang, J., Hester, A., Alqerm, I., Liu, Y. and Zhao, Y., 2018. EdgeChain: An edge-IoT framework and prototype based on blockchain and smart contracts. *IEEE Internet of Things Journal*, *6*(3), pp.4719-4732.

Papadodimas, G., Palaiokrasas, G., Litke, A. and Varvarigou, T., 2018, November. Implementation of smart contracts for blockchain based IoT applications. In *2018 9th International Conference on the Network of the Future (NOF)* (pp. 60-67). IEEE.

Peters, G.W. and Panayi, E., 2016. *Understanding modern banking ledgers through blockchain technologies: Future of transaction processing and smart contracts on the internet of money* (pp. 239-278). Springer International Publishing.

Ramachandran, A. and Kantarcioglu, D., 2017. Using blockchain and smart contracts for secure data provenance management. *arXiv preprint arXiv:1709.10000*.

Rashidi Nasab, A., & Elzarka, H. (2023). Optimizing Machine Learning Algorithms for Improving Prediction of Bridge Deck Deterioration: A Case Study of Ohio Bridges. Buildings, 13(6), p. 1517. https://doi.org/10.3390/buildings13061517

Rikken, O., Janssen, M. and Kwee, Z., 2019. Governance challenges of blockchain and decentralized autonomous organizations. *Information Polity*, *24*(4), pp.397-417.

Roszko-Wójtowicz, E. and Grzelak, M.M., 2021. Multi-dimensional analysis of regional investment attractiveness in Poland. *Equilibrium. Quarterly Journal of Economics and Economic Policy*, *16*(1), pp.103-138.

Roudini, S., Murdoch, L. C., DeWolf, S., & Germanovich, L. N. (2020). Interpretation of Borehole Strain Measurements Using Surrogate Modeling-Based Optimization. In AGU Fall Meeting Abstracts (Vol. 2020, pp. H036-0008). 2020AGUFMH036.0008R. https://ui.adsabs.harvard.edu/abs/2020AGUFMH036.0008R/abstract

Rowland, M., 2022. Trade Growth in Blockchain-based Non-Fungible Token (NFT) Markets for Digital Assets. *Smart Governance*, *1*(1), pp.49-63.

Rozario, A.M. and Thomas, C., 2019. Reengineering the audit with blockchain and smart contracts. *Journal of emerging technologies in accounting*, *16*(1), pp.21-35.

Rozas, D., Tenorio-Fornés, A., Díaz-Molina, S. and Hassan, S., 2021. When ostrom meets blockchain: exploring the potentials of blockchain for commons governance. *Sage Open*, *11*(1), p.21582440211002526.

Rymarczyk, J., 2020. Technologies, opportunities and challenges of the industrial revolution 4.0: theoretical considerations. *Entrepreneurial business and economics review*, *8*(1), pp.185-198.

Sadeghi, S., Marjani, T., Hassani, A., & Moreno, J. (2022). Development of Optimal Stock Portfolio Selection Model in the Tehran Stock Exchange by Employing Markowitz Mean-Semivariance Model. *Journal of Finance Issues*, 20(1), pp. 47-71.

Sasani, F., Mousa, R., Karkehabadi, A., Dehbashi, S., & Mohammadi, A. (2023). TM-vector: A Novel Forecasting Approach for Market stock movement with a Rich Representation of Twitter and Market data. arXiv preprint arXiv:2304.02094.

Schrepel, T., 2019. Collusion by blockchain and smart contracts. *Harv. JL & Tech.*, *33*, p.117.

Sekhar, S.M., Siddesh, G.M., Kalra, S. and Anand, S., 2019. A study of use cases for smart contracts using blockchain technology. *International Journal of Information Systems and Social Change (IJISSC)*, *10*(2), pp.15-34.

Shan, S., Duan, X., Zhang, Y., Zhang, T.T. and Li, H., 2021. Research on collaborative governance of smart government based on blockchain technology: an evolutionary approach. *Discrete Dynamics in Nature and Society*, *2021*, pp.1-23.

Sharma, A., Tomar, R., Chilamkurti, N. and Kim, B.G., 2020. Blockchain based smart contracts for internet of medical things in e-healthcare. *Electronics*, *9*(10), p.1609.

Shojaei, A., Flood, I., Moud, H.I., Hatami, M. and Zhang, X., 2020. An implementation of smart contracts by integrating BIM and blockchain. In *Proceedings of the Future Technologies Conference (FTC) 2019: Volume 2* (pp. 519-527). Springer International Publishing.

Sifah, E.B., Xia, H., Cobblah, C.N.A., Xia, Q., Gao, J. and Du, X., 2020. BEMPAS: a decentralized employee performance assessment system based on blockchain for smart city governance. *IEEE Access*, *8*, pp.99528-99539.

Sillaber, C. and Waltl, B., 2017. Life cycle of smart contracts in blockchain ecosystems. *Datenschutz und Datensicherheit-DuD*, *41*(8), pp.497-500.

Singh, A., Parizi, R.M., Zhang, Q., Choo, K.K.R. and Dehghantanha, A., 2020. Blockchain smart contracts formalization: Approaches and challenges to address vulnerabilities. *Computers & Security*, *88*, p.101654.

Singh, M.P. and Chopra, A.K., 2020. Computational governance and violable contracts for blockchain applications. *Computer*, *53*(1), pp.53-62.

Sinha, D. and Roy Chowdhury, S., 2021. Blockchain-based smart contract for international business–a framework. *Journal of Global Operations and Strategic Sourcing*, *14*(1), pp.224-260.

Sotoudehnia, M., 2021. 'Making blockchain real': regulatory discourses of blockchains as a smart, civic service. *Regional Studies*, *55*(12), pp.1857-1867.

Sultana, T., Almogren, A., Akbar, M., Zuair, M., Ullah, I. and Javaid, N., 2020. Data sharing system integrating access control mechanism using blockchain-based smart contracts for IoT devices. *Applied Sciences*, *10*(2), p.488.

Sun, J., Yan, J. and Zhang, K.Z., 2016. Blockchain-based sharing services: What blockchain technology can contribute to smart cities. *Financial Innovation*, *2*(1), pp.1-9.

Tehranian, K. (2023). Can Machine Learning Catch Economic Recessions Using Economic and Market Sentiments?. arXiv preprint arXiv:2308.16200. https://doi.org/10.48550/arXiv.2308.16200

Tehranian, K. (2023). Monetary Policy & Stock Market. arXiv preprint arXiv:2305.13930. https://doi.org/10.48550/arXiv.2305.13930

Temte, M.N., 2019. Blockchain challenges traditional contract law: Just how smart are smart contracts. *Wyo. L. Rev.*, *19*, p.87.

Triana Casallas, J.A., Cueva Lovelle, J.M. and Rodríguez Molano, J.I., 2020. Smart contracts with blockchain in the public sector. *International Journal of Interactive Multimedia and Artificial Intelligence*.

Uriarte, R.B., De Nicola, R. and Kritikos, K., 2018, December. Towards distributed sla management with smart contracts and blockchain. In *2018 IEEE International Conference on Cloud Computing Technology and Science (CloudCom)* (pp. 266-271). IEEE.

Uriarte, R.B., Zhou, H., Kritikos, K., Shi, Z., Zhao, Z. and De Nicola, R., 2021. Distributed service-level agreement management with smart contracts and blockchain. *Concurrency and Computation: Practice and Experience*, *33*(14), p.e5800.

Viano, C., Avanzo, S., Cerutti, M., Cordero, A., Schifanella, C. and Boella, G., 2022. Blockchain tools for socio-economic interactions in local communities. *Policy and Society*, *41*(3), pp.373-385.

Wallace, S. and Lăzăroiu, G., 2021. Predictive control algorithms, real-world connected vehicle data, and smart mobility technologies in intelligent transportation planning and engineering. *Contemporary Readings in Law and Social Justice*, *13*(2), pp.79-92.

Wang, S., Ouyang, L., Yuan, Y., Ni, X., Han, X. and Wang, F.Y., 2019. Blockchain-enabled smart contracts: architecture, applications, and future trends. *IEEE Transactions on Systems, Man, and Cybernetics: Systems*, *49*(11), pp.2266-2277.

Watanabe, H., Fujimura, S., Nakadaira, A., Miyazaki, Y., Akutsu, A. and Kishigami, J., 2016, January. Blockchain contract: Securing a blockchain applied to smart contracts. In *2016 IEEE international conference on consumer electronics (ICCE)* (pp. 467-468). IEEE.

Watson, R., 2022. Tradeable digital assets, immersive extended reality technologies, and blockchain-based virtual worlds in the metaverse economy. *Smart Governance*, *1*(1), pp.7-20.

Wong, P.F., Chia, F.C., Kiu, M.S. and Lou, E.C., 2022. Potential integration of blockchain technology into smart sustainable city (SSC) developments: a systematic review. *Smart and Sustainable Built Environment*, *11*(3), pp.559-574.

Wright, C. and Serguieva, A., 2017, December. Sustainable blockchain-enabled services: Smart

contracts. In *2017 IEEE International Conference on Big Data (Big Data)* (pp. 4255-4264). IEEE.

Xuan, S., Zheng, L., Chung, I., Wang, W., Man, D., Du, X., Yang, W. and Guizani, M., 2020. An incentive mechanism for data sharing based on blockchain with smart contracts. *Computers & Electrical Engineering*, *83*, p.106587.

Yakymova, L., Novotná, A., Kuz, V. and Tamándl, L., 2022. Measuring industry digital transformation with a composite indicator: A case study of the utility industry.

Zachariadis, M., Hileman, G. and Scott, S.V., 2019. Governance and control in distributed ledgers: Understanding the challenges facing blockchain technology in financial services. *Information and Organization*, *29*(2), pp.105-117.

Zhang, L., Zhang, Z., Wang, W., Jin, Z., Su, Y. and Chen, H., 2021. Research on a covert communication model realized by using smart contracts in blockchain environment. *IEEE Systems Journal*, *16*(2), pp.2822-2833.

\*\*\*